# A Wideband Composite Sequence Impedance Model for Evaluation of Interactions in Unbalanced Power-Electronic-Based Power Systems

Zhi Liu, *Student Member, IEEE,* Chengxi Liu\*, *Senior Member, IEEE*, Jiangbei Han, *Member, IEEE,* Rui Qiu, *Student Member, IEEE,* Mingyuan Liu, *Student Member, IEEE*

*Abstract*—This paper proposes a wideband composite sequence impedance model (WCSIM)-based analysis method to evaluate the interactions in power-electronic-based power systems subjected to unbalanced grid faults or with unbalanced loads. The WCSIM-based method intuitively assesses the impact of the small-signal interconnection among the positive-, negative-, and zero-sequence circuits on the interaction stability of unbalanced power systems. The effectiveness of this method is demonstrated using a permanent magnet synchronous generator-based weak grid system under a single-line-to-ground fault (SLGF). Frequency scanning results and controller hardware-in-loop tests validate both the correctness of the WCSIM and the effectiveness of the WCSIM-based analysis method.

## I. Introduction

WITH the rising penetration of power electronic devices, interaction issues in power systems become more pronounced, drawing widespread attention [1]. Unstable interaction can induce oscillations, leading to equipment tripping and even system collapse. Although various methods have been proposed to analyze and mitigate wideband oscillations, they are generally limited to balanced systems [2]. For power systems subjected to unbalanced grid faults or with unbalanced loads, the interactions become more complex, which necessitates further investigation.

Recent studies have explored interaction issues in wind power systems during unbalanced grid conditions. These studies have assessed the impact of the phase-locked loop (PLL) [3], allocation of active versus reactive currents [4], and the coupling among positive- and negative-sequence components in wind turbine generators (WTGs) [2], [5] on interaction stability. However, they primarily focus on the modeling of WTG during unbalanced grid conditions and evaluating the role of WTG controllers in interaction stability, with little attention given to the unbalanced power grid.

Unbalanced grid conditions can induce a small-signal interconnection among the positive-, negative-, and zero-sequence circuits, which potentially affects the interaction stability. It is well known that the sequence interconnection technique enables the steady-state analysis in traditional power systems without power electronic devices during unbalanced grid faults [6]-[7]. This motivates an interesting question: can this technique be applied to evaluate the wideband interactions in power-electronic-based power systems during unbalanced grid conditions?

This letter proposes a novel wideband composite sequence impedance model (WCSIM)-based method to evaluate the interactions in power-electronic-based power systems during unbalanced grid conditions. Section II presents the derivation of the WCSIM and the WCSIM-based analysis results, followed by frequency scanning results. Subsequently, Section III shows the controller hardware-in-loop (CHIL) tests. Finally, Section IV concludes the letter.

## II. The Wideband Composite Sequence Impedance Model-Based Analysis Method

### A. Description of the Test System

Fig. 1 shows a permanent magnet synchronous generator (PMSG)-based wind power plant (WPP) connected to the weak grid. 150 PMSG-based wind power generators are connected to the infinite power grid through step-up transformers, a 35-kV transmission line, and a 220-kV transmission line. Step-up transformer adopts a star connection with grounded neutral on the high-voltage side while employing a delta connection on the low-voltage side. The zero-sequence impedance of the transmission line is set to three times the positive-sequence impedance. The PMSG employs the classic vector control strategy [8], with a notch filter embedded in the PLL to suppress the negative-sequence component [4]. The detailed parameters are given in the Appendix.

### B. Derivation of the Wideband Composite Sequence Impedance Model

Suppose a single-line-to-ground fault (SLGF) occurs in phase A of the 220-kV transmission line, the fault boundary conditions in small-signal phase components can be derived as: $\dot{u}_{fa}(\omega)=R_f(s)\dot{I}_{fa}(\omega)$, $\dot{I}_{fb}(\omega)=\dot{I}_{fc}(\omega)=0$, where $\dot{u}_{fa}(\omega)$ and $\dot{I}_{fa}(\omega)$

This work was supported in part by the National Natural Science Foundation of China under Grant U22B20100, and in part by Guangdong Basic and Applied Basic Research Foundation under Grant 2023B1515250001 and Grant 2022A1515240033.

Zhi Liu, Chengxi Liu, Jiangbei Han, Rui Qiu, and Mingyuan Liu are with Hubei Engineering and Technology Research Center for AC/DC Intelligent Distribution Network, School of Electrical Engineering and Automation, Wuhan University, Wuhan, Hubei 430072, China, and also with Wuhan University Shenzhen Research Institute, Shenzhen, Guangdong 51800, China (email: zhiliu@whu.edu.cn; liuchengxi@whu.edu.cn; hanjiangbei@whu.edu.cn; qiurui@whu.edu.cn; liumingyuan@whu.edu.cn).

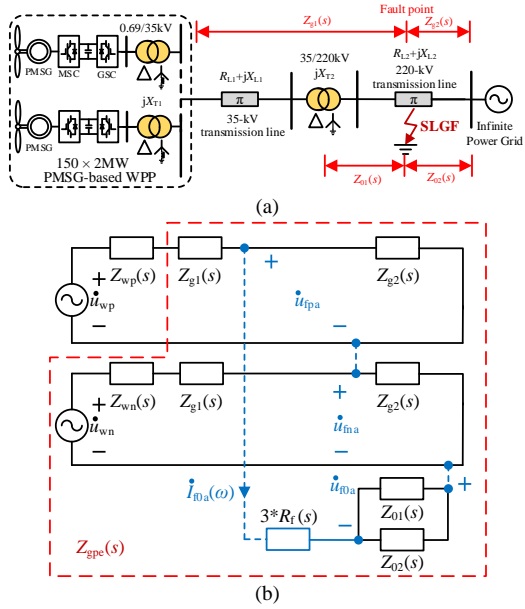

(a)

(b)

Fig. 1. PMSG-based WPP connected to the weak grid during SLGF. (a) The simplified structure. (b) The wideband composite sequence impedance model.

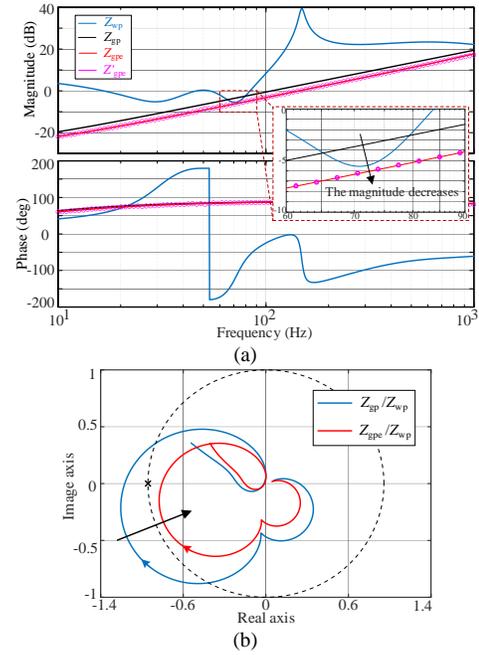

(a)

(b)

Fig. 2. (a) Bode diagrams of the theoretical and measured impedances of the WPP and the power grid. (b) Nyquist diagrams with and without SSSI

represent the small-signal components of the phase-A voltage and current at the fault point, $\dot{I}_{fb}(\omega)$ and $\dot{I}_{fc}(\omega)$ represent the small-signal components of the phase-B and phase-C of the fault branch current. Based on the symmetrical component theory [2], the fault boundary conditions can also be expressed in small-signal sequence components:

$$\begin{cases} \dot{u}_{fpa}(\omega) + \dot{u}_{fna}(\omega) + \dot{u}_{f0a}(\omega) = R_f(s)\dot{I}_{fa}(\omega) \\ \dot{I}_{fpa}(\omega) = \dot{I}_{fna}(\omega) = \dot{I}_{f0a}(\omega) = \dot{I}_{fa}(\omega)/3 \end{cases} \quad (1)$$

where $\dot{u}_{fpa}(\omega)$, $\dot{u}_{fna}(\omega)$, and $\dot{u}_{f0a}(\omega)$ represent the small-signal components of the phase-A positive-, negative-, and zero-sequence voltage at the fault point, respectively; $\dot{I}_{fpa}(\omega)$, $\dot{I}_{fna}(\omega)$, and $\dot{I}_{f0a}(\omega)$ represent the small-signal components of the phase-A positive-, negative-, and zero-sequence fault current, respectively; $\dot{u}_f(\omega)$ and $\dot{I}_f(\omega)$ represent the small-signal components of the fault voltage and current, respectively.

According to (1), this paper presents a novel wideband composite sequence impedance model (WCSIM), as shown in Fig. 1(b). The symbols $Z_{wp}(s)$ and $Z_{wn}(s)$ denote the positive- and negative-sequence WPP impedance, with their expressions given in [4], [9]; $Z_{gp}(s)$ and $Z_{gn}(s)$ denote the positive- and negative-sequence grid impedance; $Z_{01}(s)$ and $Z_{02}(s)$ denote the WPP-side and grid-side zero-sequence impedance of the 220-kV transmission line; $Z_{g1}(s)$ and $Z_{g2}(s)$ denote the WPP-side and grid-side positive-/negative-sequence grid impedance; $R_f(s)$ denotes the fault resistance. Note that the proposed WCSIM is also effective for systems with unbalanced loads, as the "fault boundary condition in small-signal components" can be satisfied by replacing the fault resistance with the small-signal impedance of unbalanced loads.

As shown in Fig. 1(b), the positive-, negative-, and zero-sequence circuits are interconnected to each other through the fault branch in phase A, rendering the phase-A quantities of the three sequence networks mutually dependent. Note that although the WCSIM is formulated for phase-A quantities, the dynamics of phase B and phase C can be considered identical to those in phase A, due to the definition of sequence components. This property will be proved in Section II-C.

As analyzed above, the dynamic behaviors of positive-, negative-, and zero-sequence components should be investigated jointly. This implies that *when evaluating the dynamics of any particular sequence circuit, it is necessary to take the impact of small-signal sequence interconnection (SSSI) into account*. For instance, in positive-sequence stability analysis, the negative- and zero-sequence impedance should be connected in parallel with the positive-sequence grid impedance, as denoted by the red dashed line in Fig. 1(b), to derive the equivalent positive-sequence grid impedance $Z_{gpe}(s)$.

$$Z_{gpe}(s) = Z_{g1}(s) + \frac{Z_{g2}(s) \cdot (Z_{ne}(s) + 3 \cdot R_f(s) + Z_{0e}(s))}{Z_{g2}(s) + Z_{ne}(s) + 3 \cdot R_f(s) + Z_{0e}(s)} \quad (2)$$

where $Z_{ne}(s)$ and $Z_{0e}(s)$ represent the equivalent negative- and zero-sequence impedance, respectively.

$$Z_{ne}(s) = \frac{(Z_{wn}(s) + Z_{g1}(s)) \cdot Z_{g2}(s)}{Z_{wn}(s) + Z_{g1}(s) + Z_{g2}(s)} \quad (3)$$

$$Z_{0e}(s) = \frac{Z_{01}(s) \cdot Z_{02}(s)}{Z_{01}(s) + Z_{02}(s)} \quad (4)$$

Then, the interaction between the WPP and the power grid can be evaluated by calculating the loop gain $Z_{gpe}(s)/Z_{wp}(s)$. The stability of the negative-sequence circuit can be analyzed using the same procedure, which will not be represented.

C. *The WCSIM-Based Analysis*

To validate the correctness of the proposed WCSIM, the frequency scanning method [10] is applied. Assuming that the SLGF occurs at 10% of the 220-kV line length from the WPP side, the theoretical and measured positive-sequence impedance curves of the power grid are presented in Fig. 2(a). Nyquist curves, considering and not considering SSSI, are given in Fig. 2(b).

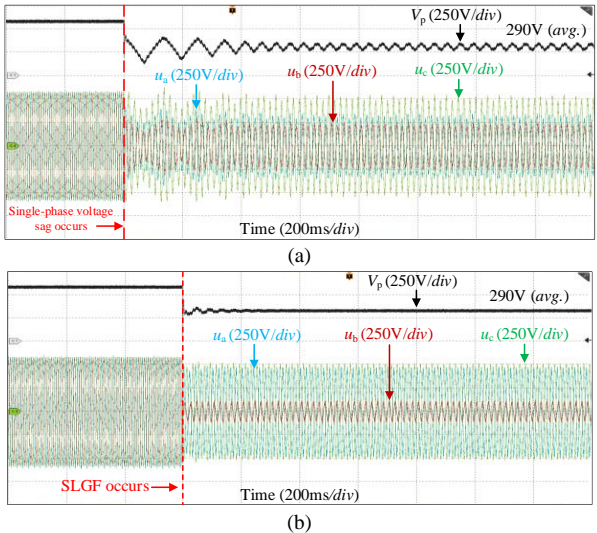

Fig. 3. The experimental waveforms. (a) Without SSSI. (b) With SSSI.

Fig. 2(a) shows that the SSSI reduces the magnitude of the grid impedance across the entire frequency range, while the phase remains unchanged. The measured grid impedance $Z'_{gpe}(s)$ matches the theoretical result in both magnitude and phase, validating the proposed WCSIM. In addition, the inset indicates that the PMSG and grid impedance curves, which previously intersected, no longer do so with SSSI, thereby increasing the damping of the super-synchronous interaction mode. The Nyquist curve in Fig. 2(b) further shows that the curve shifts from enclosing (-1, j0) to not encircling it, indicating the system becomes stable with SSSI.

## III. CONTROLLER HARDWARE-IN-LOOP TESTS

To verify the effectiveness of the proposed WCSIM-based analysis method, the test system is investigated in a CHIL platform [1]. In addition, the cases with only the single-phase voltage sag and with an actual SLGF are tested to emphasize the stability impact induced by SSSI. The experimental results are shown in Fig. 3.

As shown in Fig. 3, the PCC voltage oscillates under the single-phase voltage sag but remains stable under SLGF. Notably, the positive-sequence PCC voltage $V_p$ in both cases remains the same (290 V). This phenomenon is consistent with the theoretical analysis in Section II-C, thereby verifying the effectiveness of the proposed WCSIM-based analysis method.

## IV. CONCLUSION

This letter proposes a WCSIM-based analysis method to evaluate the interactions in power-electronic-based power systems during unbalanced grid conditions. The WCSIM differs from the traditional composite sequence network in terms of functionality and validity, as the WCSIM is valid over a wide frequency range but is only effective for evaluating wideband interactions due to its dependence on the specific equilibrium points. In addition, due to its small-signal framework, the WCSIM applies not only to power systems subjected to unbalanced faults with real fault resistance, but also to power systems with unbalanced loads characterized by "virtual fault impedance". The related work will be investigated in the future. Frequency scanning results and CHIL tests of PMSG-based WPP connected to a weak grid during SLGF verify the effectiveness of the proposed method.

## V. APPENDIX A

TABLE I
PARAMETERS OF THE EQUIVALENT TRANSMISSION LINE (BASE CAPACITY = 300MVA)

| Symbols | Value [p.u] | Symbols | Value [p.u] |
|---|---|---|---|
| $R_{L1}$, $X_{L1}$ | 0.0058+j0.058 | $X_{T2}$ | 0.03 |
| $R_{L2}$, $X_{L2}$ | 0.041+j0.41 | $R_f$ | 0.02 |

TABLE II
PARAMETERS OF PMSG-BASED WIND TURBINE GENERATORS (BASE CAPACITY = 2 MVA)

| Symbols | Definition | Value |
|---|---|---|
| $k_{pc}$, $k_{ic}$ | The proportional and integral coefficients of the current controller | 0.015, 3 |
| $k_{pp}$, $k_{ip}$ | The proportional and integral coefficients of the phase-locked loop | 0.48, 43 |
| $\omega_N$, $\zeta_N$ | Center frequency and the damping ratio of the notch filter in the phase-locked loop | 200π, 1 |
| $\omega_N$ | The rated angular frequency of the VSG [rad/s] | 100π |
| $L_f$, $C_f$, $R_{cf}$ | The inductance, capacitance, and resistance of the filter | 3mH, 50uF, 1.5Ω |
| $I_{dr}$, $I_{qr}$ | The active and reactive reference current [A] | 2368, 0 |
| $K_f$, $K_d$ | The feedforward coefficient of the d-q current and the grid voltage | 0.00157, 0.001667 |
| $V_{dc}$ | The DC voltage of the PMSG [V] | 690 |
| $X_{T1}$ | The reactance of the transformer T1 [p.u.] | 0.01 |